# Atomic Order in Non-Equilibrium Silicon-Germanium-Tin Semiconductors

S. Mukherjee[1], N. Kodali[1], D. Isheim[2], S. Wirths[3], J. M. Hartmann[4], D. Buca[3], D. N. Seidman[2], and O. Moutanabbir[1,*]

[1] *Department of Engineering Physics, École Polytechnique de Montréal, Montréal, C. P. 6079, Succ. Centre-Ville, Montréal, Québec H3C 3A7, Canada.*

[2] *Department of Materials Science and Engineering and Northwestern University Center for Atom-Probe Tomography, Northwestern University, Evanston, IL 60208-3108, USA.*

[3] *Peter Grünberg Institute 9 and JARA - FIT, Forschungszentrum Juelich, Juelich 52425, Germany.*

[4] *CEA, LETI, Minatec Campus, 17 rue des Martyrs, Grenoble 38054, France.*

The precise knowledge of the atomic order in monocrystalline alloys is fundamental to understand and predict their physical properties. With this perspective, we utilized laser-assisted atom probe tomography to investigate the three-dimensional distribution of atoms in non-equilibrium epitaxial Sn-rich group IV SiGeSn ternary semiconductors. Different atom probe statistical analysis tools including frequency distribution analysis, partial radial distribution functions, and nearest neighbor analysis were employed in order to evaluate and compare the behavior of the three elements to their spatial distributions in an ideal solid solution. This atomistic-level analysis provided clear evidence of an unexpected repulsive interaction between Sn and Si leading to the deviation of Si atoms from the theoretical random distribution. This departure from an ideal solid solution is supported by first principal calculations and attributed to the tendency of the system to reduce its mixing enthalpy throughout the layer-by-layer growth process.

The assumption that the arrangement of atoms within the crystal lattice is perfectly random is a broadly used approximation to establish the physical properties of semiconductor alloys. This approximation allows one to estimate rather accurately certain thermodynamic as well as material parameters like the excess enthalpy of formation, Vegard-like lattice parameters, and band gaps that are smaller than the composition weighted average (optical bowing). However, it has been proposed that some ternary semiconductors can deviate from this assumed perfect solid solution. Indeed, both calculations and experiments suggested the presence of local atomic order in certain III-V alloys [1–8]. This phenomenon manifests itself when at least one of the elements forming the alloy preferentially occupies or avoids specific lattice sites. This induces short-range order in the lattice with an impact on the basic properties of the alloyed semiconductors [3–7].

The recent progress in developing Sn-rich group IV (SiGeSn) ternary semiconductors and their integration in a variety of low dimensional systems and devices have revived the interest in elucidating the atomistic-level properties of monocrystalline alloys [9–20]. Interestingly, unlike III-V semiconductors, achieving a direct bandgap in $Si_xGe_{1-x-y}Sn_y$ requires a sizable incorporation of Sn ($>$ 10 at. %), which is significantly higher than the equilibrium solubility ($<$1 at.%). Understanding the atomic structure of these metastable alloys is therefore imperative for implementing predictive models to describe their basic properties. With this perspective, we present a first study of the atomic order in $Si_xGe_{1-x-y}Sn_y$ alloys (x and y in $0.04 - 0.19$ and $0.02 - 0.12$ range, respectively). We employed Atom Probe Tomography (APT) which allows atomistic level investigations [21–24] and statistical tools to analyze the three-dimensional (3-D) distributions of the three elements. This analysis unraveled an unexpected repulsive interaction between Sn and Si leading to a deviation of Si atoms from the theoretical random distribution.

The samples investigated in this work were grown using a metal cold-wall reduced pressure chemical vapor deposition (CVD) [10,25,26] using $Si_2H_6$, $Ge_2H_6$, and $SnCl_4$ as precursors and a relatively low growth temperature of 350 - 475°C resulting in a normal growth rate of ~1nm/s. Further details of epitaxial growth of $Si_xGe_{1-x-y}Sn_y$ are described in the Supplementary Material (SM). Fig.1(a) shows the STEM images of the layer with highest Sn content ($Si_{0.04}Ge_{0.84}Sn_{0.12}$), confirming the pseudomorphic growth of the ternary layers without dislocations or extended defects. Fig.1(b) shows the corresponding 3-D APT reconstructed map. The details of APT analysis are given in SM along with data sets recorded from different layers (Fig.S1). For cluster analysis, iso-concentration surfaces were first defined at varying Sn and Si concentrations within the reconstructed maps. No evidence of any aggregates was found regardless of the content. The mass spectra (Fig.S2) reveal the absence of any diatomic clusters like $(^iX_2)^{+\text{ or }++}$ or $(^iX^jX)^{+\text{ or }++}$ (where X could be Si, Ge, or Sn and i, j are isotopes of X) or higher order clusters. This behavior does not qualitatively change as a function of Sn content in the composition range investigated in this work. In order to investigate the short-range atomic distribution, we performed a series of statistical analyses, namely the frequency distribution (FD) analysis, the partial radial distribution function (p-RDF) analysis, and the nearest neighbor (NN) analysis within pre-defined regions in 3-D maps. The theoretical formalism of each method is outlined in SM. Fig.2(a) displays the FD of Si, Ge, and Sn in $Si_{0.04}Ge_{0.84}Sn_{0.12}$. The coefficient of determination ($R^2$) between the observed value ($y_i$) and the binomial distribution ($f_i$) was calculated from the residual sum of squares, $S_{res} = \sum_i (y_i - f_i)^2$ and total sum of squares, $S_{tot} = \sum_i (y_i - \bar{y})^2$ and the relation $R^2 = 1 - S_{res}/S_{tot}$ with $\bar{y} = 1/n \sum_{i=1}^{n} y_i$. The figure shows the mean values of the experimental FD for Si, Ge, and Sn correspond to 4.0, 84.0 and 12.0 at.%, respectively. This agrees with the concentrations found in the proximity histogram in Fig.S1(a). Additionally, it reveals that while Ge and Sn closely follows the binomial distribution (calculated $R^2$ for Ge and Sn are 0.9999 and 0.9978 respectively), Si shows a small disagreement with binomial distribution (calculated $R^2$ for Si is 0.9679). We eliminate statistical fluctuations as a possible reason for this observed deviation because the probed volume is large enough to have a significant number of Si atoms. In Fig.2(b), we show the p-RDF of Sn and Si in $Si_{0.04}Ge_{0.84}Sn_{0.12}$ wrt. Sn, Ge, and Si. It is worthwhile to state that in all analyses carried out for different layers, Ge is always found to be random. Henceforth, we shall restrict our discussion mainly to the behavior of Si and Sn. Also noteworthy is the fact that p-RDF is meaningful only for $r \geq 0.5\ nm$. For $r < 0.5 nm$, APT data analysis program does not find any atom and generates random values for p-RDF. The following facts are evident from Fig.2(b): Si shows a negative correlation wrt. Sn and a positive correlation wrt. Ge as well as itself; Sn has its p-RDF at unity wrt. Ge and Si while it shows signs of a positive correlation wrt. itself. However, the FD and the NN distribution (later in Fig.4) do not give any indication that Sn deviates from a perfect random distribution. We therefore think that the departure from unity shown by p-RDF of Sn wrt. itself in Fig.2(b) might be due to minute long-range compositional variations of Sn across the reconstructed APT maps.

The observations in Fig.2 provide clear evidence that Si atoms are disrupted from a perfect random distribution. The negative correlation shown by Si wrt. Sn hints at the presence of a repulsive interaction between the two species. This phenomenon provides new insights into the growth kinetics of metastable alloys by chemical vapor deposition [25], where the growth conditions prevent Sn atoms from forming equilibrium aggregated phase, despite the fact that the growth of a complete monolayer takes place in ~0.1s which is a very slow process compared to the time scale of surface diffusion events. The data presented in Fig.2 suggest that the repulsive interaction between Si and Sn is resulting from these elements diffusing away from each other during the growth, justifying the negative correlation of Si wrt. Sn. Note that Si atoms diffusing away from Sn can either hop to other Ge atoms or make enough a large number of hops to other Si atoms (which are scarce). Hence, we see a positive correlation of Si wrt. Ge and also wrt. itself. Since bulk diffusion is energetically less favorable than surface diffusion, it is reasonable to conclude that the observed departure for an ideal solid solution occurs during the layer-by-layer growth. It is important to note that this phenomenon is peculiar to Sn-rich ternary alloys since extended x-ray absorption fine structure (EXAFS) investigations indicated the atomic distribution in Sn-rich strained and relaxed GeSn binary alloys to be random [27], also asserting the fact that epitaxial strain is not responsible for the observation we made in Fig 2.

Interestingly, the analysis of ternary layers with lower Sn contents ($\leq 4$ at. %) indicates that the aforementioned departure from a perfectly random alloy is either absent or too small to be detected. For instance, Fig.3(a) exhibits the FD for each element in $Si_{0.10}Ge_{0.875}Sn_{0.025}$. Noteworthy is the overlap between the observed distribution and the binomial distribution (black lines) assuming a complete random alloy ($R^2$ was calculated for Si, Ge, and Sn to be 0.9989, 0.9999, and 0.9959, respectively). Fig.3(b) shows the measured Sn and Si p-RDF within a sphere of radius 10 nm. Here, the p-RDF of Sn wrt. Sn shows the telltale signature of statistical fluctuations owing to its small concentration of only 2.5 at. %. The p-RDF fluctuates around the mean value of unity with the magnitude of these fluctuations decreases with increasing $r$. Note that the volume of the shell considered during p-RDF analysis and consequently the number of atoms which lies inside the shell increases as a function of $r^2$. Finally, the p-RDF of Sn steadies down to the value of unity. The p-RDF of Si wrt. Si and Sn are qualitatively similar, none showing any noticeable deviation from 1. The p-RDF of Sn wrt. Si, Ge and Si wrt. Ge (Fig.S3) confirm that at low Sn content, all atoms are randomly distributed within the alloy, reinforcing the results of the FD analysis.

Fig.4(a) displays the analysis of NN distribution $NN_{A-A} - k$, where 'A' is Si or Sn; $k$ is 5 (5$^{th}$ NN) or 10 (10$^{th}$ NN), for $Si_{0.04}Ge_{0.84}Sn_{0.12}$. While $NN_{Sn-Sn}$ follow the probability distribution closely, $NN_{Si-Si}$ show some peculiar features. In fact, for $k = 5$ and 10, $NN_{Si-Si}$ distributions show a shoulder that appears at a slightly lower value of $r$ than the maxima of $P_k(r, C)$. The rest of the measured $NN_{Si-Si}$ distribution essentially follows $P_k(r, C)$ but with a very minute right shift. The departure of the $NN_{Si-Si} - 5$ and $NN_{Si-Si} - 10$ from the binomial value (Fig.4(b)) provides a clear indication of a disruption in Si distribution creating local pockets where there are more Si wrt. a given Si atom. In these pockets, the Si concentration is slightly higher than the average bulk concentration making the average distance between a given Si atom and its 5$^{th}$ or 10$^{th}$ NN slightly smaller than what is expected theoretically. The minute right shift also indicates that the rest of the matrix is slightly depleted of Si making the 5$^{th}$ and 10$^{th}$ NN distance slightly larger than that in a perfect random distribution. The shoulder is obviously absent in $NN_{Si-Si} - 1$ (Fig.S4(a)) due to the fact that no atom can be located at a distance smaller than the first nearest neighbor distance. Also noteworthy is the fact that such features are absent in the $NN_{Si-Si} - 5$ and $NN_{Si-Si} - 10$ distributions at very low Sn content of 2.5 at. % (Fig.S4(b)). It must however be remembered that the observed deviation of Si in Sn-rich alloys from a perfect random atomic distribution must not be confused with the formation of aggregates [28,29].

The revelation that Si atoms in monocrystalline Sn-rich ternary alloys deviate from the behavior in an ideal solid solution is indeed surprising. We attributed this disruption in Si distribution to a repulsive interaction between the Sn and Si atoms. In order to elucidate the energetics of this phenomenon, we performed detailed DFT calculations using the Quantum Espresso code (details in SM) on a 32 atom supercell. As shown in Fig.S6, we found that a Si-Sn bond is indeed energetically not favorable requiring an additional energy of $\sim +50 - 250$ meV/unit cell as compared to the most stable configurations ($NN-3$ and $NN-4$). A similar observation was made during a recent EXAFS study on SiGeSn ternary alloys [30]. These calculations support qualitatively the hypothesized repulsive interaction between Si and Sn atoms. Here, it is important to notice that the incorporation of a large radius Sn atom in Ge lattice would create a local distortion leading to more compressive Ge regions around Sn. This seems to affect the incorporation of Si, which has a smaller radius, and perhaps prefers to incorporate in available sites far from these compressive regions. What we must remember, however, is that the growth of a metastable alloy is a kinetically controlled non-equilibrium process. Owing to limited mobility, atoms after deposition on the surface are inhibited to reach the equilibrium state within the growth time-scale. The problem therefore reduces to a surface process where one needs to evaluate how atoms behave at the surface before they become buried underneath the next growing layer. Herein, one can reasonably neglect bulk diffusion as it implies energy barriers that are significantly higher than those for surface diffusion. One can also intuitively assume that, once atoms are deposited on a surface, the system will begin to evolve to minimize the mixing enthalpy $\Delta H_{mix}$ in an effort to reduce its Gibbs free energy. This evolution is abruptly brought to an end after the next growing layer sweeps across the entire surface. Since Ge is the solvent and Si and Sn are solutes, the mixing enthalpy $\Delta H_{mix}$ is given by: $\Delta H_{mix} = H_{alloy} - xH_{Si} - (1-x-y)H_{Ge} - yH_{Sn}$. With the mole fractions and the enthalpies of the pure elements ($H_{Si}, H_{Ge}, H_{Sn}$) predetermined, $H_{alloy}$ becomes a determining parameter. $H_{alloy}$ is affected by factors like epitaxial strain, micro-strain, and chemical interaction. Micro-strain which arises when the lattice has to accommodate two atoms of dissimilar size yet maintaining a uniform lattice constant throughout the crystal. For example, with N the total number of atoms and $\Omega_{Si-Sn}$ the Si-Sn interaction parameter, the micro-strain contribution coming from Si-Sn bonds in a regular solution is given by $N\Omega_{Si-Sn}(xy)$. Unlike the epitaxial strain and micro-strain whose contribution to $H_{alloy}$ is always positive, the chemical interaction contribution ($\Delta H_{Ch}$) to $H_{alloy}$ can be positive or negative. This depends on the nature of charge transfer between two atoms forming a bond. For example, calculations showed that in ordered GaInP$_2$ the difference in electronegativity between the atoms caused charge to flow from the less ionic Ga − P bond to the more ionic In − P, giving a small positive value of $\Delta H_{Ch}$ [1]. The process of charge transfer takes place not only for bonding atoms which belong to different groups (hence different electronegativity) but also for isovalent heteropolar atoms like group IV elements. Indeed, first-principle calculations found a small positive and a small negative value of $\Delta H_{Ch}$ for SiGe and SiC respectively [31].

In summary, we performed atomic scale studies on Sn-rich metastable SiGeSn ternary alloys using APT. To investigate the randomness in the distribution of different atoms within the alloys, we implemented different statistical techniques, namely the FD, p-RDF's, and the NN distribution. Our study shows that the Si atoms deviate from a perfectly random solid solution within the alloy with large Sn content. The phenomenon is attributed to a repulsive interaction between Sn and Si, thereby inducing local disruptions in an otherwise random distribution of Si.

The DFT calculations also demonstrated that having Si and Sn atoms as nearest neighbors is indeed energetically unfavorable. These departures from an ideal solid solution shown by Si is either absent or too weak to be detected in alloys with low Sn content ($< 4$ at. %). The observed short range ordering must be taken into account for a more accurate evaluation of lattice parameter, lattice relaxation, thermodynamic parameters, band structure, and opto-electronic properties of group IV ternary semiconductors.

**Acknowledgment.** The work was supported by NSERC-Canada, Canada Research Chair, Calcul Québec, and Compute Canada. The LEAP at the Northwestern University Center for Atom-Probe Tomography (NUCAPT) was acquired and upgraded with equipment grants from the MRI program of the National Science Foundation (grant number DMR-0420532) and the DURIP program of the Office of Naval Research (grant numbers N00014-0400798, N00014-0610539, N00014-0910781). NUCAPT is supported by the NSF's MRSEC program (grant number DMR-1121262). Additional instrumentation at NUCAPT was supported by the Initiative for Sustainability and Energy at Northwestern (ISEN).

[*]Corresponding author.
oussama.moutanabbir@polymtl.ca

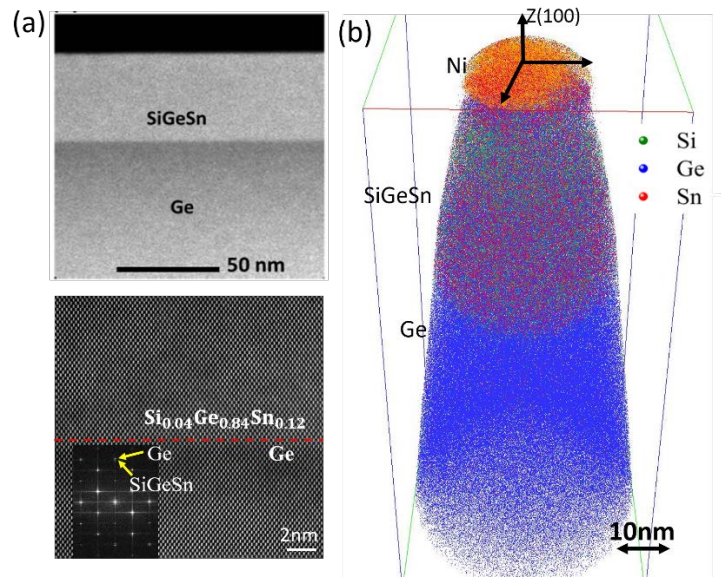

**FIG. 1.** STEM image, 3D atom-by-atom reconstruction (a) High angle annular dark field STEM (top) and high resolution STEM (bottom) images of the $Si_{0.04}Ge_{0.84}Sn_{0.12}$/Ge interface. Inset: A diffraction pattern taken from a selected region at the interface (b) 3-D reconstruction of the ternary alloy ($Si_{0.04}Ge_{0.84}Sn_{0.12}$), showing the Ni capping layer, the SiGeSn thin film and a portion of the Ge buffer layer. For the sake of clarity, only 10% of Ge atoms and 50% of Sn atoms are displayed

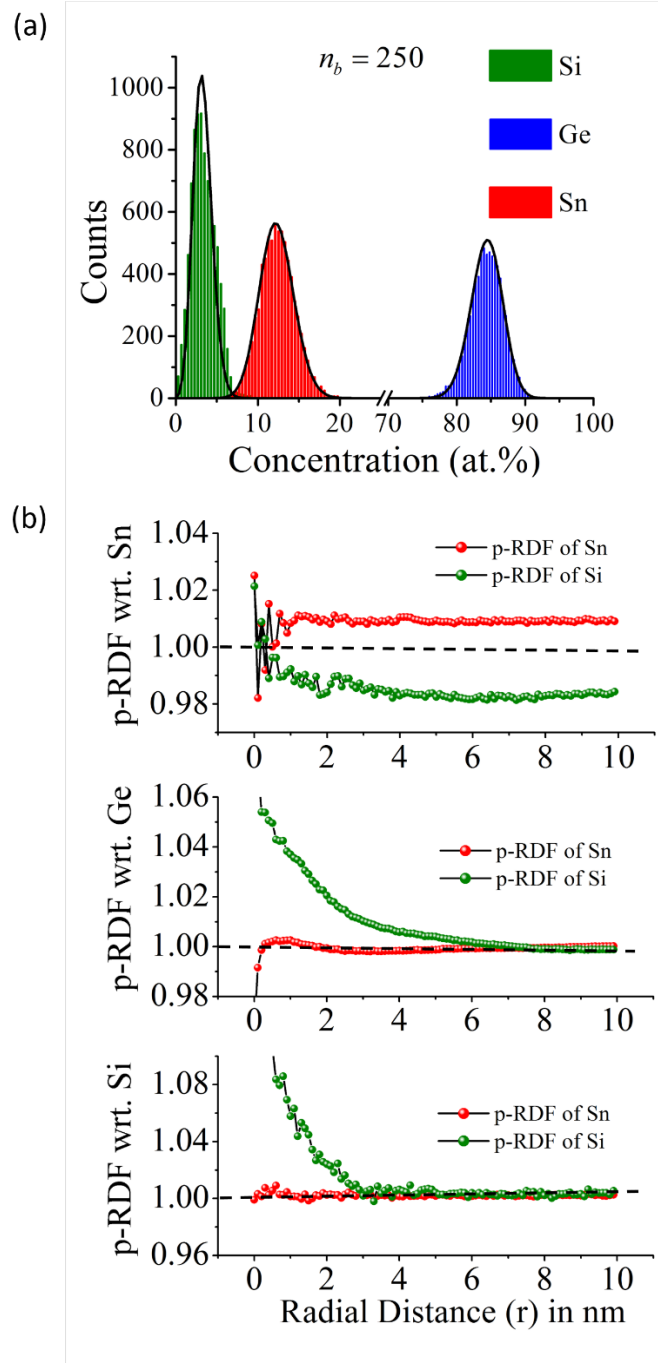

**FIG. 2**. Frequency distribution and p-RDF of the constituents in the alloy containing 12.0 at.% Sn. (a) Frequency distribution of Si (green), Ge (blue), and Sn (red) in $Si_{0.04}Ge_{0.84}Sn_{0.12}$ as determined from APT reconstruction (in histograms). The corresponding binomial distribution of these atoms are shown in black continuous lines (b) The partial radial distribution function of Sn and Si atoms with respect to Sn (top), Ge (middle), and Si (bottom) in the same sample as in (a) for r = 10nm. The IVAS computed error bars are smaller than the data symbols. The black dotted line represents $p - RDF = 1$

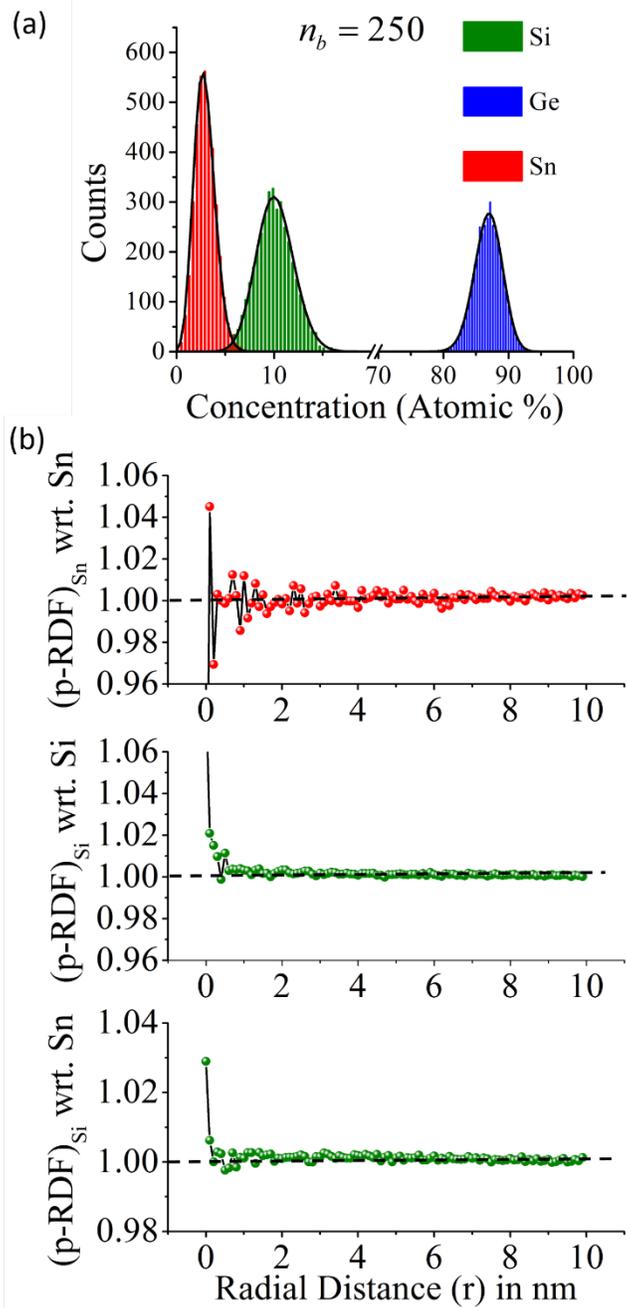

**FIG. 3**. Frequency distribution and p-RDF of the constituents in the alloy containing 2.5 at.% Sn. (a) Frequency distribution of Si (green), Ge (blue), and Sn (red) in $Si_{0.10}Ge_{0.875}Sn_{0.025}$ as determined from APT reconstruction (in histograms). The corresponding binomial distribution of these atoms are shown in black continuous lines. (b) The partial radial distribution function in the same sample as (a) for r = 10nm of Sn atoms wrt. Sn (top), Si wrt. Si (middle), and Si wrt. Ge (bottom). The IVAS computed error bars are smaller than the data symbols. The black dotted line represents $p - RDF = 1$

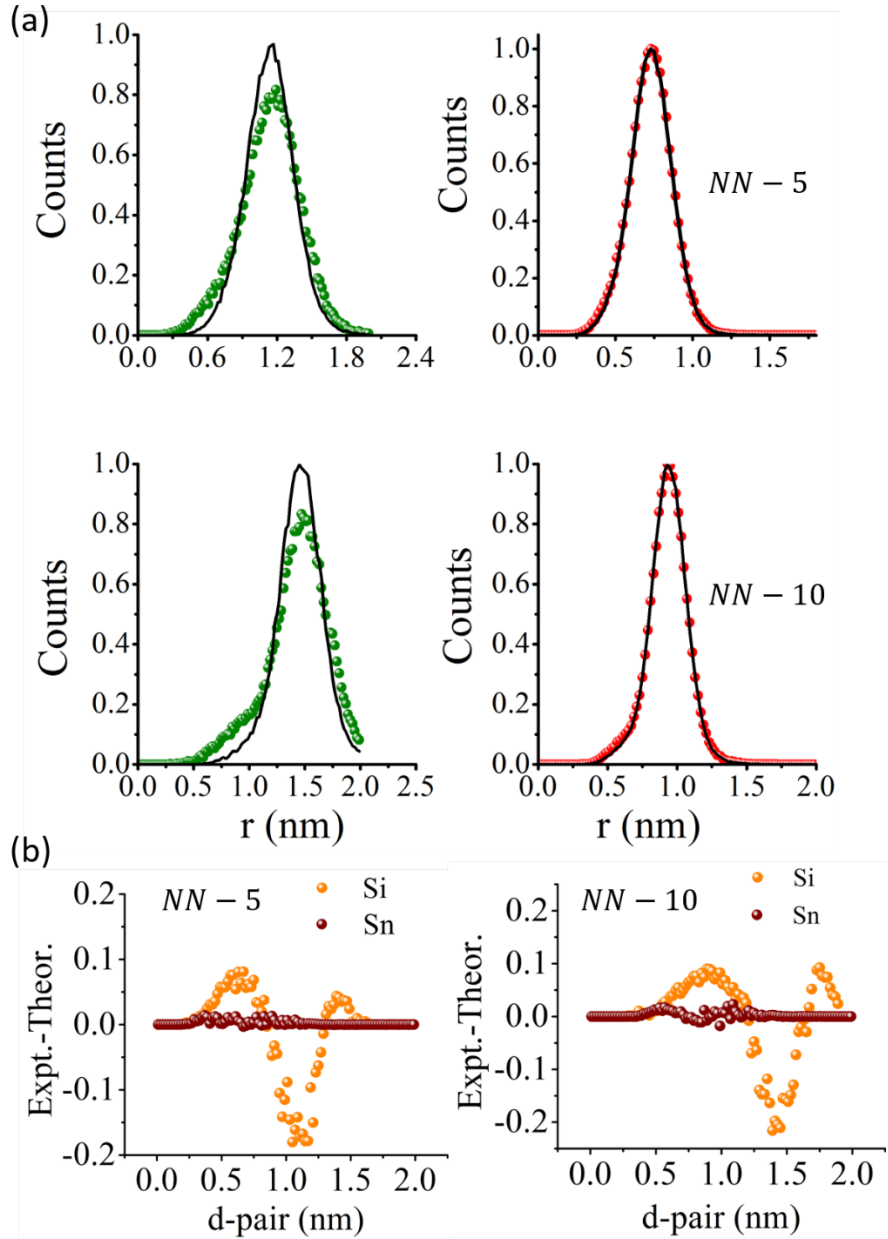

**FIG. 4**. The nearest neighbor distribution (a) Si-Si and Sn-Sn NN-5 and 10 distribution in the alloy containing 12.0 at.% Sn. The distribution as determined from APT reconstruction are shown in solid spheres Si (green) and Sn (red). The corresponding binomial distribution are shown in black continuous lines. All the data sets are normalized with respect to the theoretical probability distribution, $P_k(r, C)$. (b) Departure of the observed Si-Si and Sn-Sn NN-5 and 10 distributions from the binomial distribution.

**Supplementary Material**

# Atomic Order in Non-Equilibrium Silicon-Germanium-Tin Semiconductors

S. Mukherjee[1], N. Kodali[1], D. Isheim[2], S. Wirths[3], J. M. Hartmann[4], D. Buca[3], D. N. Seidman[2], and O. Moutanabbir[1,*]

[1] *Department of Engineering Physics, École Polytechnique de Montréal, Montréal, C. P. 6079, Succ. Centre-Ville, Montréal, Québec H3C 3A7, Canada.*

[2] *Department of Materials Science and Engineering and Northwestern University Center for Atom-Probe Tomography, Northwestern University, Evanston, IL 60208-3108, USA.*

[3] *Peter Grünberg Institute 9 and JARA - FIT, Forschungszentrum Juelich, Juelich 52425, Germany.*

[4] *CEA, LETI, Minatec Campus, 17 rue des Martyrs, Grenoble 38054, France.*

**Details of growth of the ternary alloys:** The growth was carried out on low-defect density Ge/Si(100) virtual substrates developed in Ref. [1] The thickness of the Ge layer atop the Si(001) was ∼2.7 μm. Si$_2$H$_6$, Ge$_2$H$_6$ (10% diluted in H$_2$), and SnCl$_4$ were used as precursors and N$_2$ as carrier gas. Relatively low growth temperatures (350 - 475°C) were used to avoid segregation, phase separation, and epitaxial breakdown thus ensuring the growth of high crystalline quality Si$_x$Ge$_{1-x-y}$Sn$_y$ with an Sn content above the equilibrium composition (>1at.%). The general approach to obtain these non-equilibrium alloy is to deposit the material one atomic layer at a time, under growth conditions in which the surface mobility of atoms is limited to such an extent that the nucleation of separated phases on the surface is prevented. Then, with the deposition of the next layer, these atoms become buried and freeze in a mixed state. The thickness of the investigated layers is ∼ 50 nm.

**Atom Probe Tomography Investigation.** The APT sample preparation was done using the standard lift-out technique [2]. The APT experiment was carried out using a UV laser-assisted Local Electrode Atom Probe (LEAP). Field evaporation of individual atoms was assisted by focusing a UV laser (λ=355nm), with a beam waist smaller than 5μm, on the apex of the needle-shaped specimen. The evaporation rate (ion/pulse), the laser pulse repetition-rate, and energy per pulse were 0.01, 500 kHz, and 25 pJ respectively. The base temperature and base pressure within the APT chamber were maintained at 50 K and $3.2 \times 10^{-11}$ Torr respectively. The 3-D reconstructions were performed using Cameca's IVAS program.

**Statistical Formalism.**

**i) The frequency distribution analysis:** In the frequency distribution analysis, the ROI is broken down into '$N'$' blocks each containing equal number of atoms, $n_b$. The total number of atoms of a particular element is then counted in each block and frequency at which a particular atom occurs is then compared with a binomial distribution. For example, if $n$ is the number of atoms of a particular element which is randomly distributed throughout the ROI and has a bulk normalized concentration of $C$ then its frequency of occurrence must follow the binomial distribution [3]:

$$f(n) = \frac{N'n_b!}{n!\,(n_b - n)!} C^n (1 - C)^{(n_b - n)}$$

**ii) The partial radial distribution function (p-RDF) analysis:** In (p-RDF) analysis the $k^{th}$ atom of an element X (which could be Si, Ge, or Sn) is chosen as the central atom. Then a spherical shell of radius $r$ and thickness $dr$ is defined around this central atom and the number of atoms of the $i^{th}$ element (Si, Ge, or Sn) positioned within the volume $4\pi r^2 dr$ is calculated. Next, $r$ is incremented by small steps and the number of atoms are computed after each step. This continues till $r$ reaches the pre-assigned maximum radius. If the atoms are ideally random (completely uncorrelated), then the exact number of atoms at $r$ within the shell of thickness $dr$ is the average number of atoms per unit volume times $4\pi r^2 dr$. Thus, $C_i^0 = \rho_i 4\pi r^2 dr$, where $\rho_i$ is the density of the $i^{th}$ element assuming it is randomly distributed, that is $\rho_i = N_i/V$. To quantify atomic ordering in a given material, the (p-RDF) is defined as [4]:

$$p - RDF = \frac{1}{C_i^0} \sum_{k=1}^{N_X} \frac{N_i^k(r)}{N_{tot}^k(r)}$$

with $N_i^k(r)$ as the actual number of atoms of the $i^{th}$ element within the shell of thickness $dr$ around the $k^{th}$ atom of element X at the center, $N_{tot}^k(r)$ is the total number of atoms of all atomic species within the shell around the $k^{th}$ atom of element X at the center, and $N_X$ is the total number of X atoms within the ROI. It is clear from the expression of (p-RDF) that if the $i^{th}$ element is randomly distributed or completely uncorrelated with respect to (wrt.) an element X, its (p-RDF) should be unity. (p-RDF) greater than unity symbolized positive correlation or a concentration greater than the bulk normalized concentration. Conversely, an (p-RDF) less than unity symbolized negative correlation indicating a concentration below the bulk normalized concentration.

**iii) The nearest neighbor (NN) distribution analysis:** For an ABC-type ternary alloy various combinations of NN distribution can be calculated like $NN_{A-A}$, $NN_{A-B}$, $NN_{B-C}$ and so on. Here, we have analyzed only the $NN_{A-A}$ type but for first, fifth, and tenth order. The NN distribution analysis calculates the distance between an 'A' atom and its closest neighboring 'A' atoms. This gives the first nearest neighbor distance or $NN_{A-A} - 1$. This is then extended to calculate the distance between the 'A' atom and its higher $k^{th}$ order neighbors, giving $NN_{A-A} - k$. The process is then repeated until the entire ROI is processed. For an ideal solid solution, where the atoms of an element are randomly distributed with a bulk normalized concentration of $C$, the probability of its $k^{th}$ nearest neighbor of a given atom being at a distance $r$ is given by the following distribution function [5]:

$$P_k(r, C) = \frac{3}{(k-1)!}\left(\frac{4\pi}{3}C\right)^k r^{3k-1} e^{-(4\pi/3)Cr^3}$$

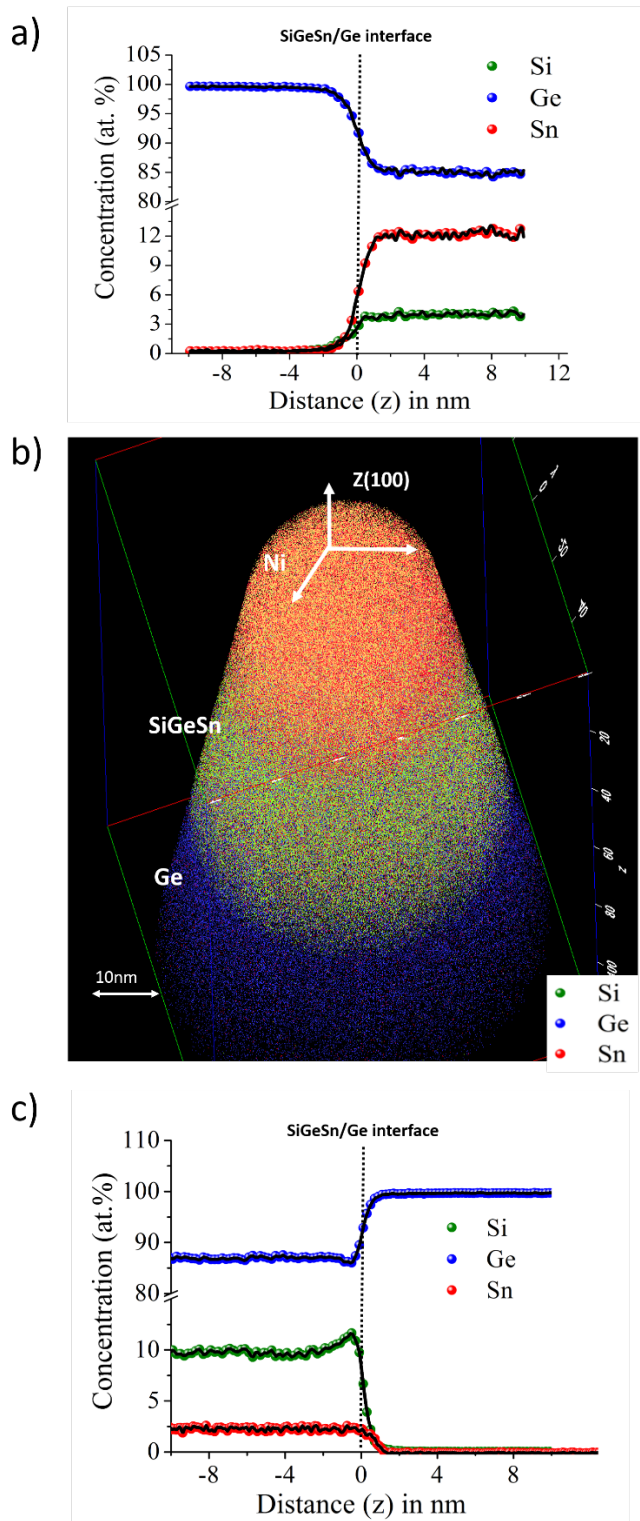

**Fig. S1.** (a) 1D proximity histogram of the alloy with high Sn content ($Si_{0.04}Ge_{0.86}Sn_{0.12}$) across the SiGeSn/Ge interface (b) 3-D reconstruction and (c) 1D proximity histogram across the SiGeSn/Ge interface of the ternary alloy with lowest Sn content ($Si_{0.04}Ge_{0.84}Sn_{0.12}$).

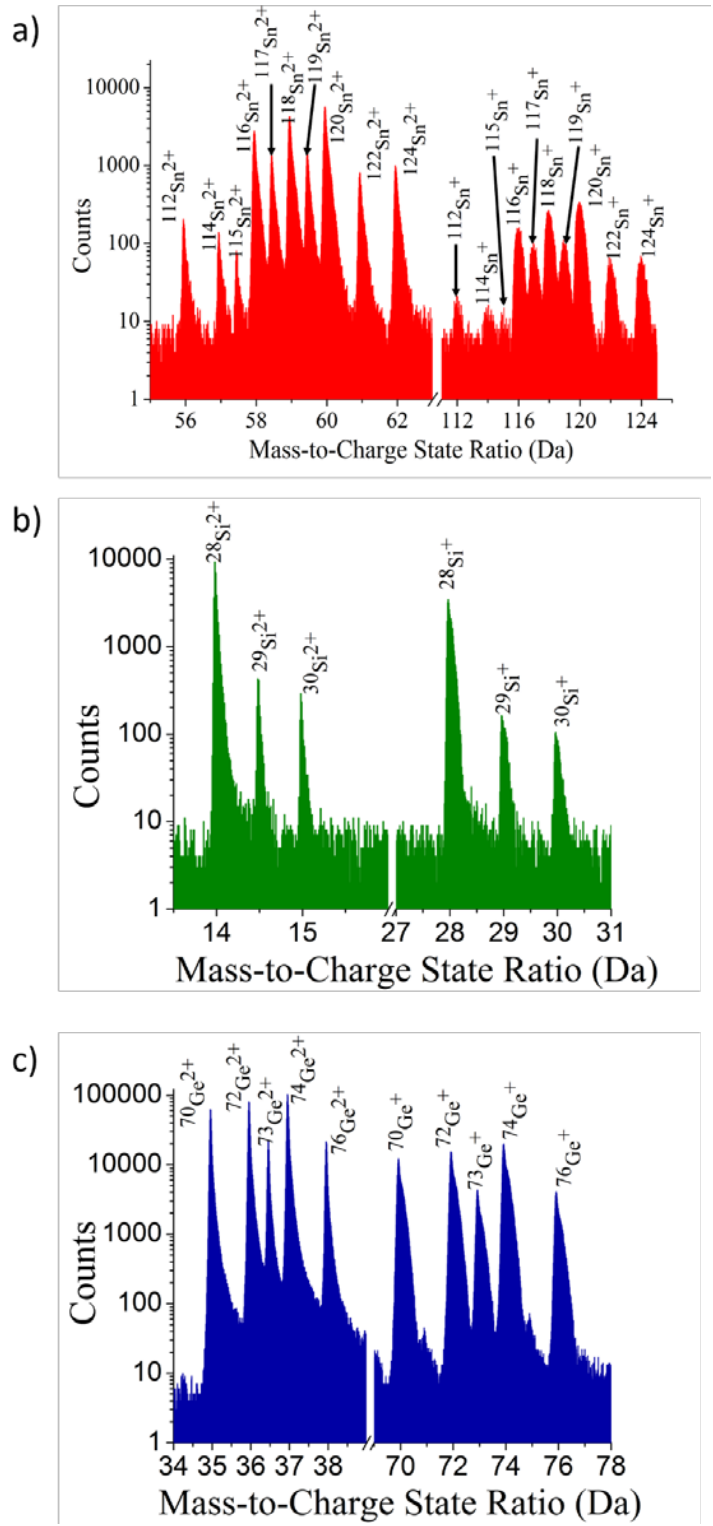

**Fig. S2.** Mass spectra of (a) Sn, (b) Si, and (c) Ge respectively in the $Si_{0.04}Ge_{0.84}Sn_{0.12}$ alloy. Both single and doubly charged states of all the isotopes are shown. The y-axis of the figures is in log scale and the x-axis has been broken to include both charge states in a single graph.

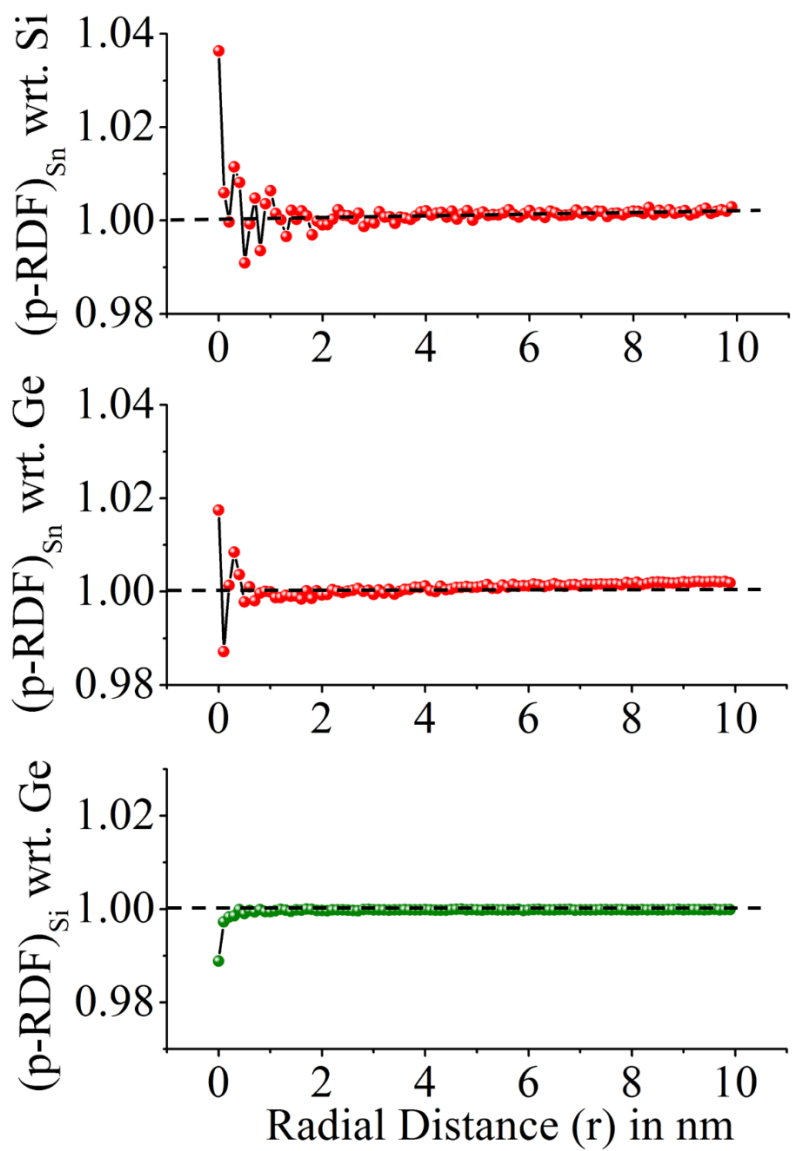

**Fig. S3**. (a) The partial radial distribution function (RDF) of Sn wrt. Si (top). Sn wrt. Ge (middle), and Si wrt. Ge (bottom) atoms in with lowest Sn content ($Si_{0.10}Ge_{0.875}Sn_{0.025}$) for r = 10nm. The IVAS computed error bars are smaller than the data symbols. The black dotted line represents p − RDF = 1

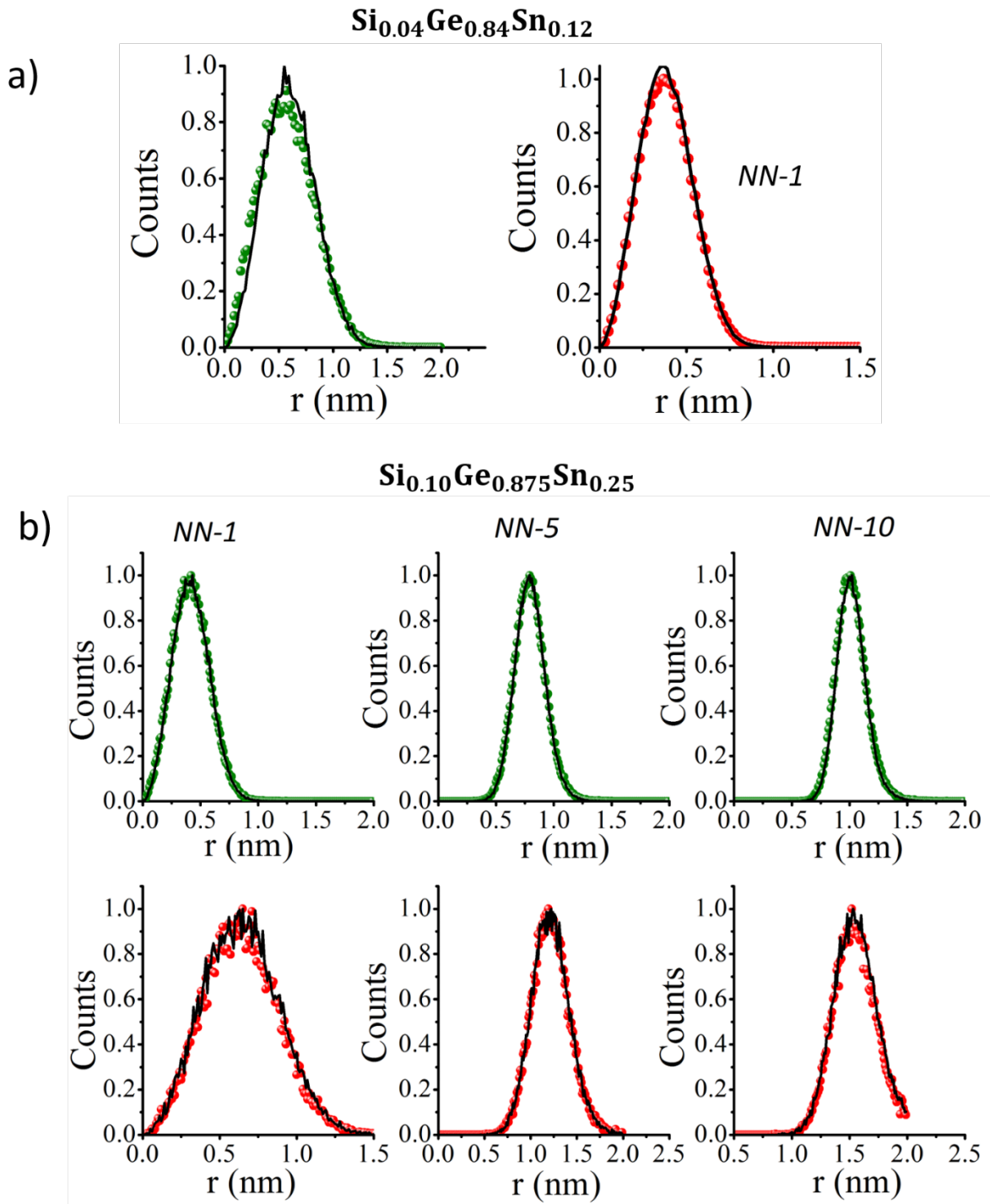

**Fig. S4**. (a) The NN-1 distribution of Si (green) and Sn (red) in the sample with high Sn content ($Si_{0.04}Ge_{0.84}Sn_{0.12}$) (b) The NN-1, 5, 10 distribution of Si (green) and Sn (red) in the sample with low Sn content ($Si_{0.10}Ge_{0.875}Sn_{0.025}$). In both the plots, the distribution determined from APT reconstruction are shown in solid spheres and the corresponding theoretical distribution considering a perfect random alloy are shown in black continuous lines.

**DFT Calculations:** The density-functional theory (DFT) calculations were using the Quantum Espresso code [6], with a plane-wave energy cutoff of 60 $R_y$ ($R_y$ is the Rydberg constant equaling 13.6 eV). The pseudopotentials used were a combination of ultrasoft [7] and PAW [8] pseudopotentials with the Perdew-Burke-Ernzerhof functional [9]. These pseudopotentials were taken from the Pslibrary project [10] and the GBRV project [11]. The reciprocal space has been sampled using a Monkhorst-Pack mesh [12] of 9 K-points. We have analyzed the (100) layer of Ge lattice by considering a supercell with 32 Ge atoms (4×4 Germanium unit cells with interlayer spacing of about 17Å) and 32 H atoms which were added to avoid surface reconstruction and remove dangling bonds. Finally, we have replaced Ge atoms with Sn and/or Si atoms as necessary to study the interactions between Sn and Si atoms in Ge lattice. We performed variable cell relaxation calculations considering the $x$ component of the first lattice vector, the $y$ component of the second lattice vector and all the atomic positions as variables. The BFGS quasi-Newton algorithm was used for both Ion and Cell dynamics.

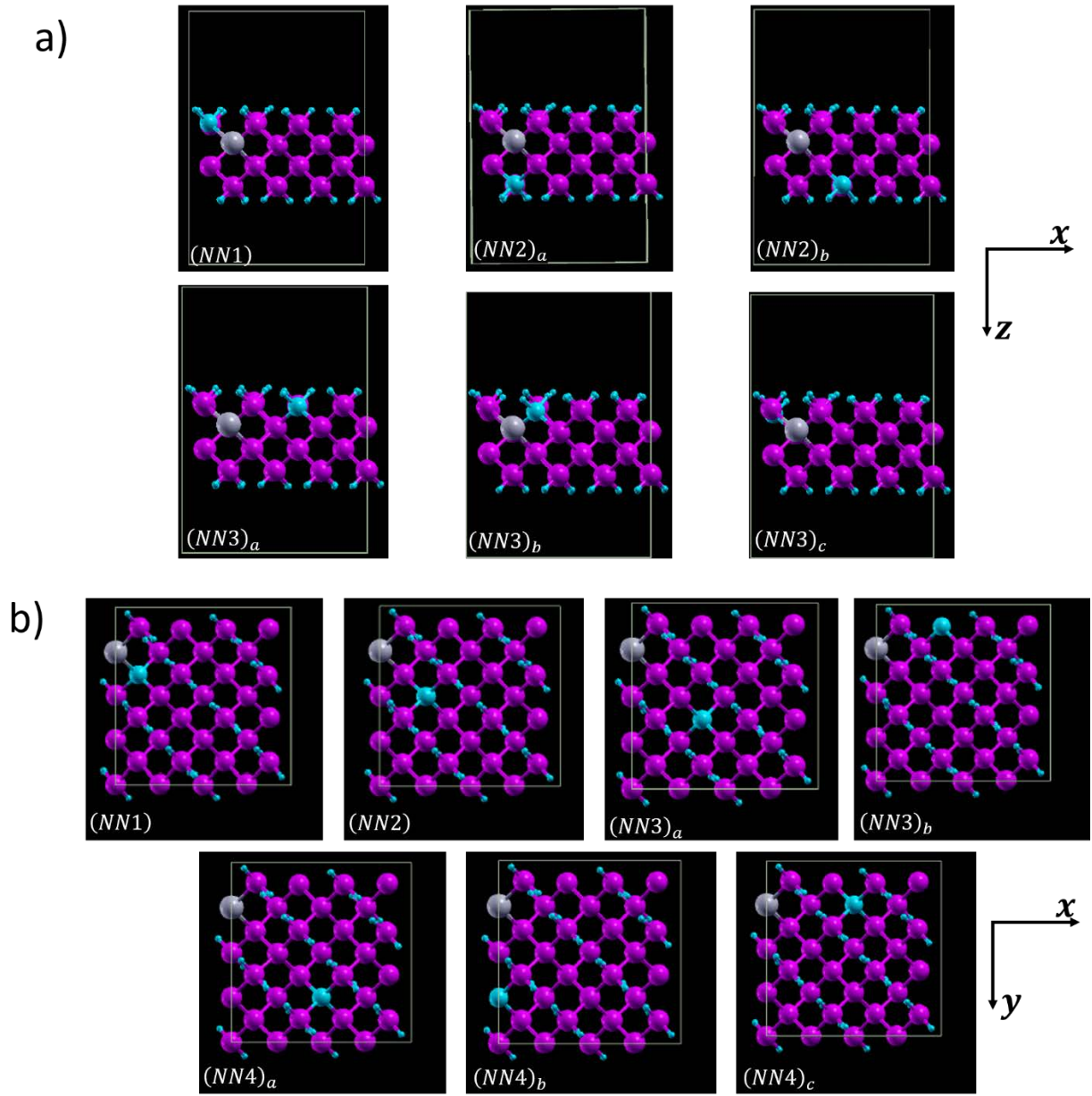

**Fig. S5**. The Si-Sn NN configurations considered for the DFT calculations. The Ge atoms are shown in purple, Sn in grey and Si in light blue. The Sn atom is always placed at the second layer from top. (a) Case 1: The Si atoms are placed in the outermost layers. (b) Case 2: The Si atoms are placed in the inner two layers.

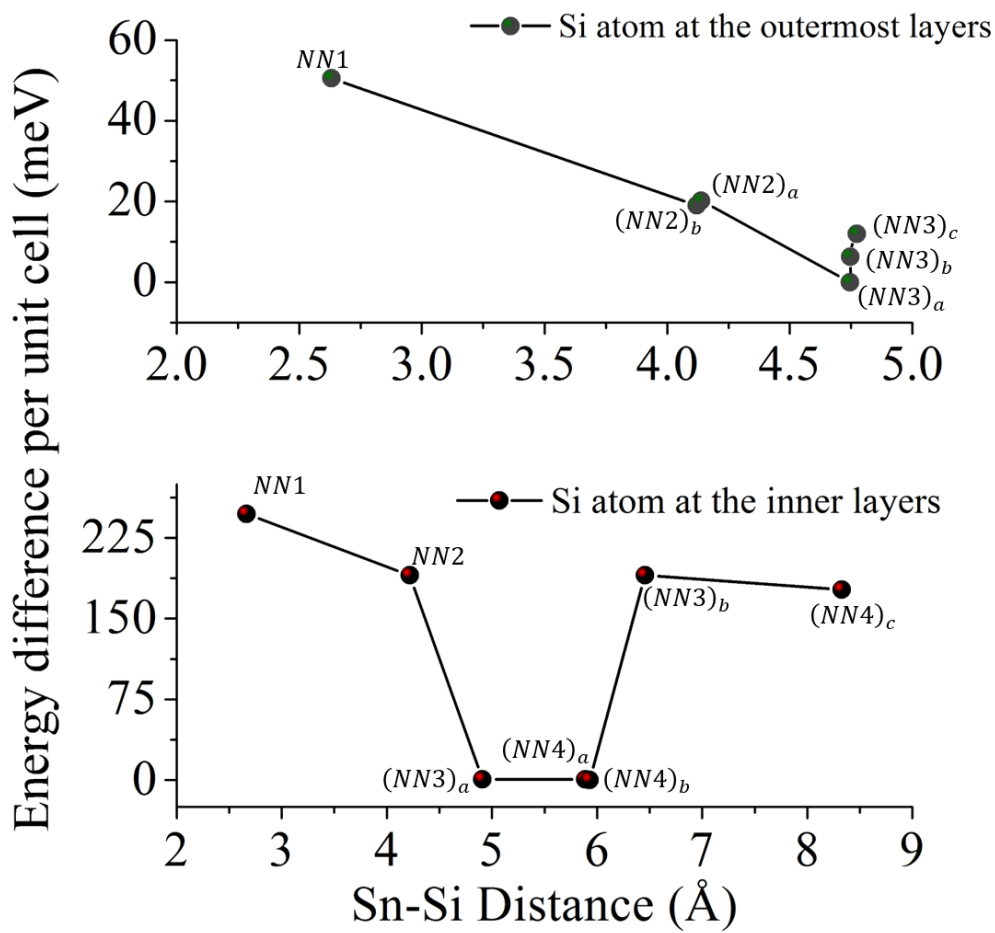

**Fig. S6**. DFT results plotting the energy difference per unit cell as a function of the Sn-Si distance. The Sn atom was always placed at the second layer from top. In the top graph, the Si atom are located only in the outermost layers of the $4 \times 4$ unit cell while in the bottom graph the Si atom was placed in the inner layers of the unit cell. The NN configurations has also been mentioned alongside each data point. The lowest energy configurations were the $(NN3)_a$ in the top graph and the $(NN4)_b$ in the bottom graph. The energy difference has been plotted relative to these lowest energy configurations. The solid black line is for the guide to the eye.

There are obviously a very large number of configurations in which the Sn and Si atoms can be arranged within the ternary alloy lattice. Herein, without losing the sense of generality, we present the result obtained for a few cases. The Sn atom is always placed at the second layer from the top and the Si atom is either placed at the outermost layers of the unit cell (case 1, Fig. S5(a)) or at the inner layers of the unit cell (case 2, Fig. S5(b)). The energy of the unit cell for both the cases for different Sn-Si distance was calculated and the difference in energy per unit cell relative to the lowest energy configuration was plotted as a function of Si-Sn distance in Fig. S6 (case 1 on top and case 2 at the bottom). The nearest neighbor configurations are mentioned alongside each data point in Fig. S6 and corresponds to the schematic representation shown in Fig. S5(a) and (b). The lowest energy was found to be the $(NN3)_a$ configuration for case1 and the $(NN4)_b$ configuration for case 2. Interestingly, we found that a Si-Sn bond (NN-1 configuration) is indeed energetically not favorable in all cases (~+50meV/unit cell for case 1) and (~+250meV/unit cell for case 2), supporting the hypothesized repulsive interaction between Si and Sn atoms.